\documentclass[nocopyright,hyphenate,balance,nofoot]{asmejour}

\hypersetup{%
	pdftitle={A Bayesian Approach for Inferring Sea Ice Loads},
	pdfkeywords={Sea Ice, Bayesian Inference, Uncertainty Quantification},
	pdfauthor={M. Parno, T. Hodgdon, B. West, D. O'Connor, A. Song},
}

\usepackage{lineno}
\usepackage{amsmath}
\usepackage{xspace}
\usepackage{algorithm}
\usepackage[noend]{algpseudocode}
\usepackage{mathtools}
\usepackage{siunitx}
\usepackage{bm}
\usepackage{tikz}
\usepackage[normalem]{ulem}
\usepackage{setspace}
\usepackage{color}
\usepackage{xcolor}

\usetikzlibrary{shapes,arrows,positioning,fit,calc}

\DeclareMathOperator{\tr}{tr}

\DeclareGraphicsExtensions{.png,.pdf}
\graphicspath{{Figures/}}

\newcommand{\reviewchange}[1]{#1}

\newcommand{\load}[1]{\ensuremath{f}}
\newcommand{\strain}[1]{\ensuremath{\epsilon_{obs}}}
\newcommand{\post}[1]{\ensuremath{P(f|\epsilon_{obs})}}
\newcommand{\likeli}[1]{\ensuremath{P(\epsilon_{obs}|f)}}
\newcommand{\prior}[1]{\ensuremath{P(f)}}

\newcounter{mycomment}

\JourName{Applied Mechanics}
\PaperYear

\begin{document}


\SetAuthorBlock{Matthew Parno\CorrespondingAuthor}{%
US Army Engineer Research and Development Center, \\
Hanover, NH 03755 USA \\
email: matthew.d.parno@erdc.dren.mil
}

\SetAuthorBlock{Taylor Hodgdon}{%
US Army Engineer Research and Development Center, \\
Hanover, NH 03755 USA\\
taylor.s.hodgdon@erdc.dren.mil
}

\SetAuthorBlock{Brendan West}{%
US Army Engineer Research and Development Center, \\
Hanover, NH 03755 USA \\
email: brendan.a.west@erdc.dren.mil
}

\SetAuthorBlock{Devin O'Connor}{%
US Army Engineer Research and Development Center, \\
Hanover, NH 03755 USA \\
email: devin.t.oconnor@erdc.dren.mil
}

\SetAuthorBlock{Arnold Song}{%
Dartmouth College
Hanover, NH 03755 USA \\
email: arnold.j.song@dartmouth.edu
}

\title{A Bayesian Approach for Inferring Sea Ice Loads}


\begin{abstract}
The Earth's climate is rapidly changing and some of the most drastic changes can be seen in the Arctic, where sea ice extent has diminished considerably in recent years. As the Arctic climate continues to change, gathering in situ sea ice measurements is increasingly important for understanding the complex evolution of the Arctic ice pack. To date, observations of ice stresses in the Arctic have been spatially and temporally sparse. We propose a measurement framework that would instrument existing sea ice buoys with strain gauges. This measurement framework uses a Bayesian inference approach to infer ice loads acting on the buoy from a set of strain gauge measurements. To test our framework, strain measurements were collected from an experiment where a buoy was frozen into ice that was subsequently compressed to simulate convergent sea ice conditions. A linear elastic finite element model was used to describe the response of the deformable buoy to mechanical loading, allowing us to link the observed strain on the buoy interior to the applied load on the buoy exterior.
 The approach presented in this paper presents an instrumentation framework that could use existing buoy platforms as in situ sensors of internal stresses in the ice pack.
\end{abstract}

\date{\today}
\maketitle

\section{Introduction}\label{sec:intro}


In the 1980s, multiyear sea ice accounted for approximately 50-60\% of the total Arctic sea ice area, but in recent years has only accounted for about 15\% \citep{comiso2012, screen2010central, jeffries2013arctic, carmack2015toward, strong2013arctic}.  This dramatic reduction in multiyear ice is part of a significant change in the character of Arctic sea ice, which continues to thin and diminish.  As a result, commercial and military activity in the Arctic is likely to increase, necessitating a better understanding of sea ice dynamics and better tools for predicting ice behavior \citep{aksenov2017future, ho2010implications, melia2016sea, pizzolato2016influence, stroeve2014changes}.

Predictions of ice behavior are typically generated from computational models of the ice dynamics.  Both continuum and discrete element methods rely on rheological models to describe the generally nonlinear relationship between stress and strain within the ice.  These relationships are often derived for long temporal and spatial scales relevant to climate scale models, where the rheology represents a homogenization of finescale processes like fracture and ridging.  However, on length scales relevant to operations ($\approx$1 km), the ice behavior is more heterogeneous \citep{schreyer2006elastic, hutchings2011spatial, marsan2004scale, rampal2008scaling} and the behavior of individual features, like ridges and leads, become more important.  In order to better understand the processes that drive ridging and fracture, it is  necessary to understand and model the local rheology of the ice and to characterize the ice dynamics over short temporal and spatial scales.  To better our understanding, in situ observations of stress and strain in the ice are needed.   

Previous field experiments involving in situ stress observations have been spatially localized and have only employed a small number of sensors for a limited amount of time \citep{johnson1985kadluk, comfort1990field, tucker1992stress, coon1993, richter1998, hata2015}.  We argue that improving ice rheologies and mechanical process models will require distributed internal stress measurements that continuously monitor the ice stress state.  Larger areas of observation and longer monitoring periods are needed to constrain the rheological parameters and to better characterize processes with data that are currently spatially or temporally sparse.  Our proposed methodology can help with this by providing a framework that is inexpensive, easy to deploy, and can potentially leverage existing sensing platforms, e.g., buoys that are part of the International Arctic Buoy Programme \citep{zweng2018inventory}.

Strain is relatively straightforward to measure because the ice deformation can be measured directly.  As \cite{cox1983stress} points out however, it is not possible to measure stress directly.  Instead, the deformation or strain of another object placed in the ice with known, typically elastic, mechanical behavior must be employed.  Using this idea to indirectly measure internal stresses has a long history, starting in earnest with the early works of \cite{Metge1975, Templeton1979, cox1983stress}.  Indeed, sensors based on the original design of \cite{johnson1980osi} are still produced commercially \citep{geokon}. Our proposed approach uses the same fundamental concept as these early efforts: measuring the deformation of an elastic body frozen in the ice.  However, whereas as previous systems were special sensors designed specifically for generating point estimates of the strain, we propose to use larger existing buoy platforms and to provide a more comprehensive view of internal ice stress by characterizing the spatially varying pressure field acting between the ice and the buoy.  More precisely, we propose a Bayesian inference framework to infer the ice stress applied to the outer wall of a hollow, steel-walled buoy embedded in an ice sheet given a handful of strain observation on the buoy interior.  This is made possible by relating the exterior stresses and observed strain with a high fidelity structural finite element model of the buoy.  Rigorous approaches to inverse problems, like the one proposed here, are quite powerful and widely used in diverse fields such as medical imaging, geologic exploration, glaciology, and structural health monitoring to estimate quantities that are difficult, expensive, or impossible to measure directly \citep{borcea2002electrical, schulz2010hybrid, morlighem2010, steiner2008time, vanik2000bayesian}.   However, we have seen limited use of such techniques in the development of in situ ice sensors.

 It is important to note that we do not model the ice itself but instead couple a structural model of the buoy to the ice via spatially distributed traction boundary conditions.  Of course, this problem is ill-posed; the traction boundary condition is in theory a function over position, i.e., an infinite dimensional quantity, and therefore cannot be constrained by a finite number of data points.  Even after discretizing the traction, we need to infer a parameterized field with many degrees of freedom that cannot be completely constrained by the finite number of available strain observations. In fact, there are many traction fields that will match the strain observations equally well.   In addition to being ill-posed, observational noise introduces further uncertainty into the problem.  Additional knowledge of the traction field (e.g., smoothness or anticipated values) is therefore \reviewchange{needed} to help overcome the ill-posedness of the problem.  Both the ill-posedness and observation noise are naturally accounted for probabilistically in the Bayesian setting employed below.


To assess the utility of our framework, we use observations from a compression test performed at the US Army Cold Regions Research and Engineering Laboratory (CRREL) in the Spring of 2018.  Section \ref{sec:methods} describes this experiment, the measurement setup, the Finite Element Method (FEM) forward model, and the Bayesian inference framework.  In Section \ref{sec:results} results are presented for both synthetic data and actual observations from the CRREL experiment.  Key patterns in the resulting traction field are related to events witnessed during the experiment. We provide discussion about the strengths and weaknesses with the inference approach in Section \ref{sec:discussion}. We conclude with remarks in Section \ref{sec:conclusion}.



\section{Methods}\label{sec:methods}
\subsection{Methods Overview}
To test the concept of using a buoy instrumented with strain gauges to estimate sea ice internal stresses, we subjected a thin-walled steel buoy to compression within a laboratory grown saline ice sheet.  Computationally, we used a CAD representation of the buoy, provided by the University of Washington Applied Physics Laboratory, to develop a structural finite element model of the buoy.  This model allowed us to describe the connection between applied traction fields and observed strain and to ultimately define an inverse problem for characterizing the traction field given actual observations.  Section \ref{subsec:experiment} describes the experimental facility and setup in more detail. Section \ref{subsec:FE} then describes the specifics of our finite element forward model.  Finally, Section \ref{subsec:bayes_method} formulates a Bayesian inference problem for connecting this model with experimental data in order to characterize the distribution over various loading scenarios.

\subsection{Physical experiment}
\label{subsec:experiment}
The ice compression experiment was conducted in the Geophysical Research Facility located at the Cold Regions Research and Engineering Laboratory in Hanover, NH.  The facility is composed of a large outdoor tank (dimensions: width 6.7 m, length 18.3 m, depth 2.3 m) and a retractable roof fitted with a refrigeration system designed to facilitate the growth of ice for subsequent mechanical testing. A compressive load was applied to the ice using a hydraulic ram composed of a set of three hydraulic pistons that can together apply a maximum load of 14 MPa. 

\begin{figure}[t]
	\centering
	\includegraphics[width=3.25in]{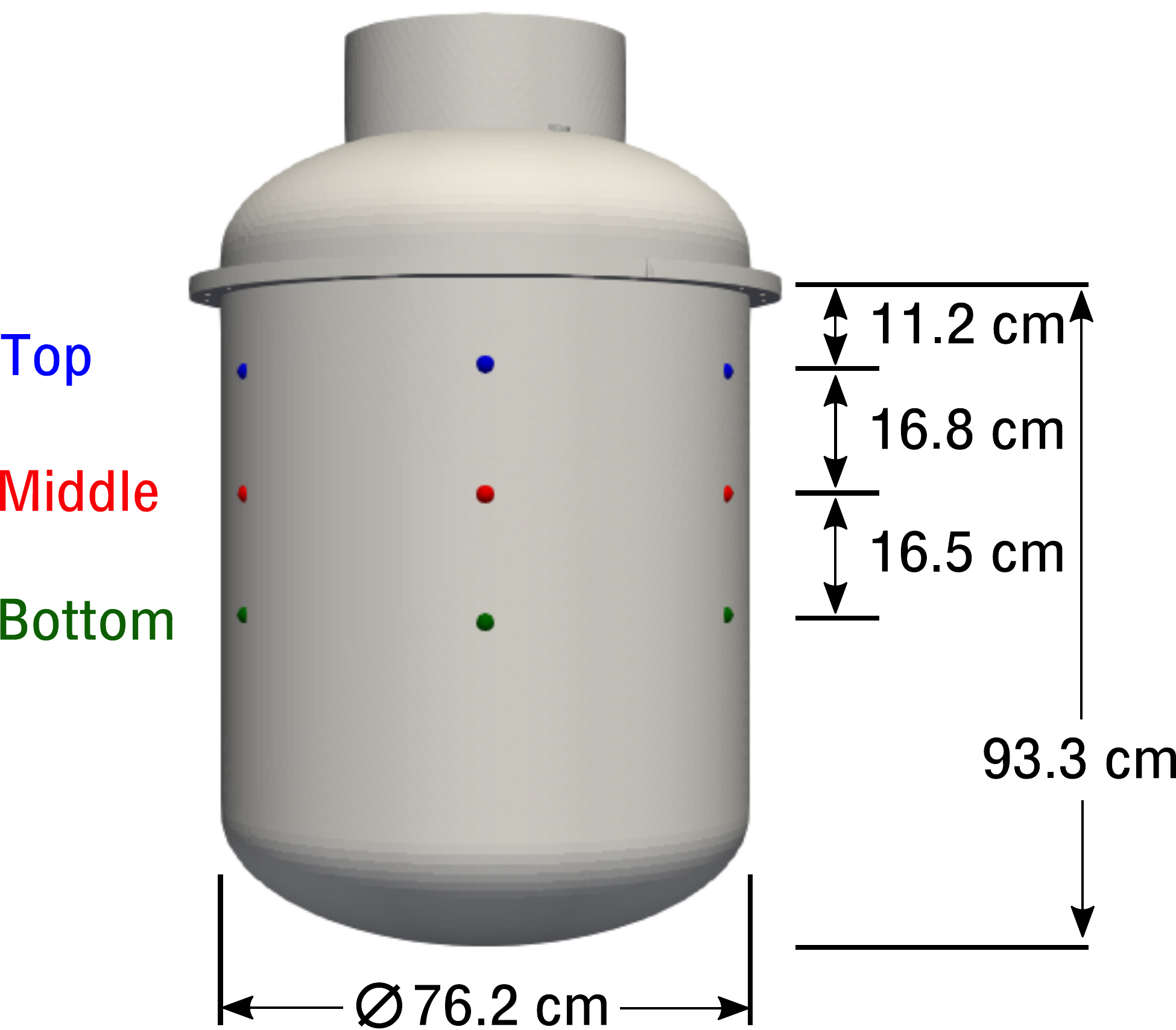}
	\caption{Dimensions of the buoy and strain gauge locations from the compression experiment.  Strain gauges are placed in three rows (top, middle, and bottom) around the circumference of the buoy at \ang{60} intervals.  Each point location has both a vertical and horizontal strain gauge.  The colors correspond to the strain results in Section \ref{sec:results}.}
	\label{fig:gauge_locations}
\end{figure}

As shown in Figure \ref{fig:gauge_locations}, the buoy was constructed from a cylindrical pressure tank with an outer diameter measuring 76.2 cm and vertical length of 116.8 cm. The pressure vessel was laterally sliced 23.5 cm from the top of the buoy and outfitted with a bolted flange to provide access to the strain measurement system on the interior of the buoy. The buoy was instrumented with 36 strain gauges arranged in pairs that measured vertical and horizontal strain at 18 locations. The strain gauges were placed in three rings around the buoy and vertically spaced 16.5 cm apart, beginning with the top ring of 12 gauges located 34.7 cm from the buoy top (11.2 cm below the bottom of the buoy cap flange), a middle ring located at 51.2 cm from the top (27.7 cm below the bottom of the buoy cap flange), and the bottom ring located 67.7 cm from the top (44.2 cm below the bottom of the buoy cap flange). The gauges were uniformly distributed radially in $\ang{60}$ intervals (see Figure \ref{fig:gauge_locations}). 

To simulate sea ice, the initial tank salinity was set to 27 ppt, which is slightly lower than typical Arctic sea surface salinity ($\sim$29-32 ppt) \citep{brucker2015}. This salinity level was selected to account for the brine rejection that typically occurs during sea ice formation. The ice growth began on December 19, 2017, and continued until February 12, 2018, reaching a thickness of 23 cm. At this stage of ice growth, the buoy was placed within a square section that was cut into the ice. The ice growing process resumed until reaching the target thickness of 50 cm. During the \reviewchange{ ice-growing} process, the test basin was covered and refrigerated until the experiment took place to promote accelerated ice growth and to limit melting during warm, sunny periods. 

\begin{figure}[t]
	\centering
	\includegraphics[width=3.25in]{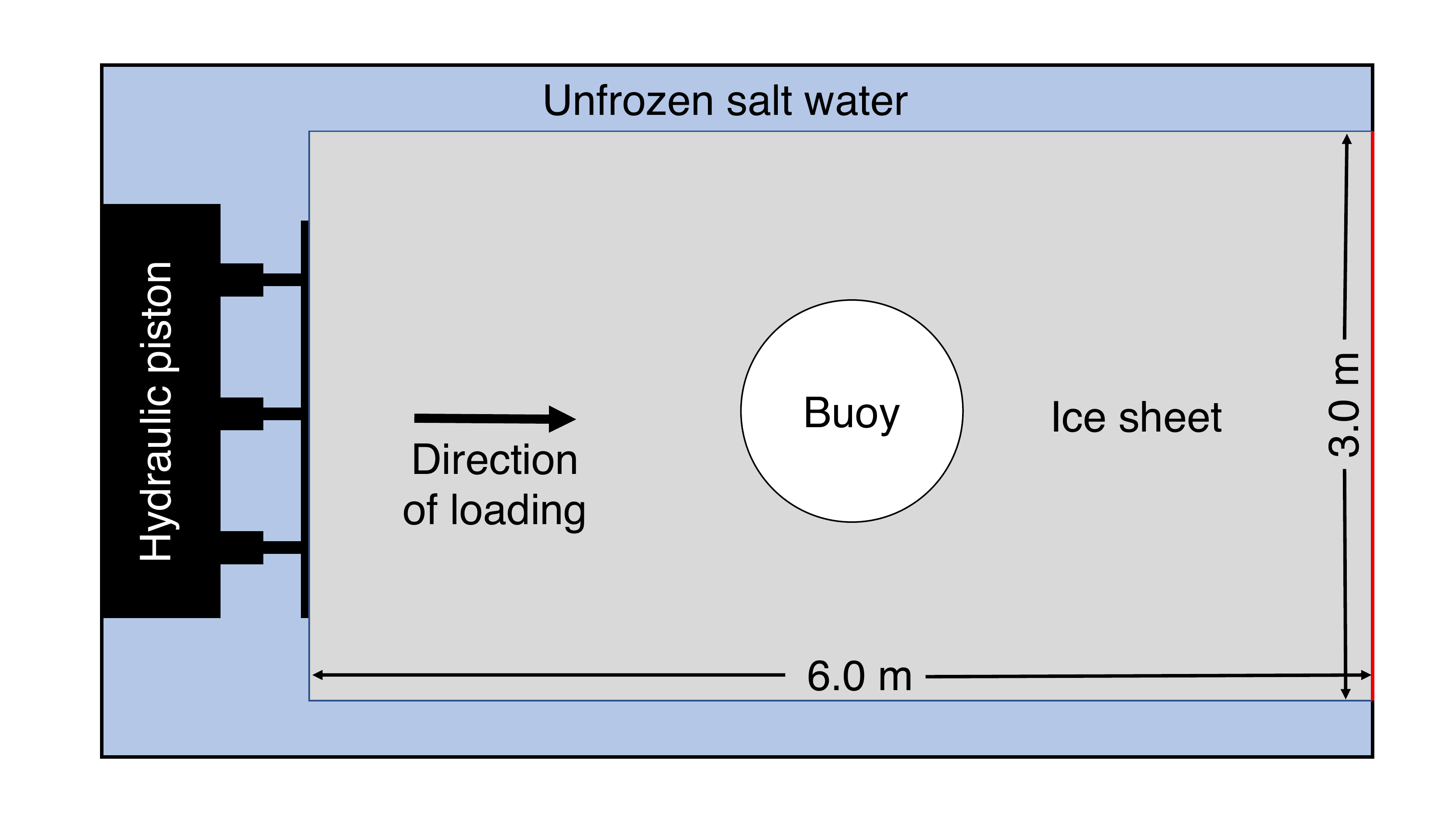}
	\caption{Schematic diagram of the setup for the buoy compression experiment. The thick black lines indicate the test basin boundary and the red line indicates the boundary where the ice was frozen to the test basin wall.}
	\label{fig:setup}
\end{figure}

The ice compression experiment was conducted on April 11, 2018, which was a partly cloudy day with an outdoor temperature of \ang{9}C.  The configuration and load condition resembles a uniaxial compression test, illustrated in Figure \ref{fig:setup}. The sides of the ice sheet parallel to the loading direction were free of any confinement. The side adjacent to the hydraulic press had a free slip boundary condition, but the opposite side was fixed.  For the test, the hydraulic ram load was gradually increased throughout the experiment until reaching a peak value of approximately 14 MPa. The experiment lasted approximately 4 minutes with strain measurements recorded every 0.5 seconds.

\subsection{Finite element model}
\label{subsec:FE}
\begin{figure}[t]
	\centering
	\includegraphics[width=3.25in]{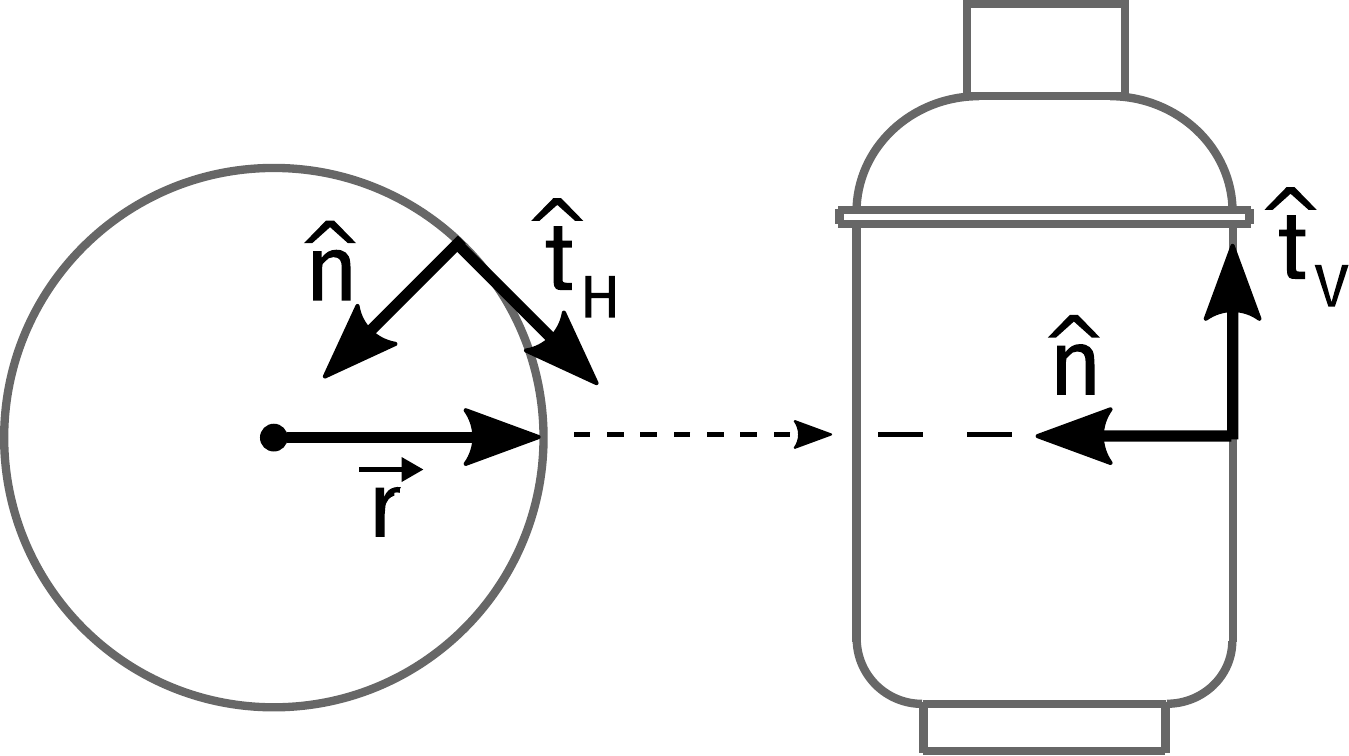}
	\caption{Schematic diagram of the unit vectors (horizontal: $\hat{t}_{H}$, vertical: $\hat{t}_{V}$, and normal: $\hat{n}$) used to project the pressure results between Cartesian coordinates and the buoy reference coordinates.  The left image is a cross-sectional view looking from the top of the buoy down, such that $\hat{t}_{V}$ is directed out of the page toward the reader.  The right image is a cross-sectional view of the buoy rotated \ang{90} about the dashed line. $\vec{r}$ is the radius of the buoy.}
	\label{fig:force_diagram}
\end{figure}

For each strain gauge location in Figure \ref{fig:gauge_locations} in the horizontal (i.e., in the x-y plane tangent to the buoy surface) and vertical strain was observed on the interior of the buoy.  To relate this to the tractions acting on the exterior of the buoy, we define a model predicting the buoy's deformation given a particular traction field.   This is done by solving the balance of linear momentum equation with the assumption that inertial effects are negligible.  This is a reasonable assumption since the loading rate of the ice on the buoy in the experiment is small.   With this assumption, the balance of linear momentum is given by
\begin{linenomath*}
\begin{equation}
	\nabla \cdot \sigma(x) + b(x) = 0,
\label{eq:stress_mod}
\end{equation}
\end{linenomath*}
where $\sigma(x)$ is the Cauchy stress tensor and $b(x)$ is the body force vector on the buoy. Assuming a linear elastic constitutive model, the stress-strain response follows Hooke's law
\begin{linenomath*}
\begin{equation}
	\sigma(x) = 2 \mu \varepsilon_{mod}(x) + \lambda \tr(\varepsilon_{mod}(x)) I,
\label{eq:constitutive_law}
\end{equation}
\end{linenomath*}
where $\mu$ and $\lambda$ are Lam\'e parameters and $\varepsilon_{mod}$ is the model strain, which is defined as (assuming small strain)
\begin{linenomath*}
\begin{equation}
	\varepsilon_{mod}(x) = \frac{1}{2} \big(\nabla u(x) + (\nabla u(x))^\intercal \big),
\label{eq:strain_mod}
\end{equation}
\end{linenomath*}
where $u$ is displacement.  We prescribe a Young's modulus of 200 GPa and a Poisson's ratio of 0.3, which correspond to the material properties of the steel buoy used in the experiment. We then derive the Galerkin weak form of Eq.\eqref{eq:stress_mod} and arrive at the discretized finite element matrix equations
\begin{linenomath*}
\begin{equation}
	Kd=f,
\label{eq:FEM_matrix_eq}
\end{equation}
\end{linenomath*}
where $K$ is the stiffness matrix, $d$ is the nodal displacement vector, and $f$ is the external nodal force vector acting on the buoy surface $\Gamma$. For the purposes of inference, we only need to know predicted values of strain at the physical strain gauge locations on the buoy.   Using the finite element basis functions that define the displacement $u$ with the definition of strain in Eq.\eqref{eq:strain_mod}, we are able to write the strain at the observation locations as 
\begin{linenomath*}
\begin{equation}
	\varepsilon_{mod} = BK^{-1}f,
\label{eq:strain}
\end{equation}
\end{linenomath*}
where $B$ is a matrix that contains the derivatives of the finite element shape functions. The size of the B matrix is constructed such that only the shape functions for the finite elements containing the location of the physical strain gauges are included.  It is a sparse matrix with $33$ rows, one for each strain gauge, and $203,661$ columns, one for each degree of freedom in the finite element mesh. The nodal force vector $f$ can be further broken down into
\begin{linenomath*}
\begin{equation}
	f = DAp
\label{eq:strain2}
\end{equation}
\end{linenomath*}
where $D$ is a matrix of shape function products integrated over the external surface $\Gamma$, similar to a mass matrix, $p$ is a nodal pressure vector, and $A$ is a matrix that rotates and stamps the pressure vector from the buoy coordinate system, shown in Figure \ref{fig:force_diagram}, into the global coordinate system.  The matrix $A$ has $203,661$ rows, one for each displacement degree of freedom in the mesh, and $35,073$ columns, one for each degree of freedom we wish to infer.  The pressure vector $p$ is composed of normal, $p_{N}$, and tangential components in the horizontal and vertical directions, $p_{H}$ and $p_{V}$, respectively. 
The matrix $A$ transforms the results between Cartesian global coordinates and the buoy coordinates through the expression
\begin{linenomath*}
\begin{equation}
        \begin{bmatrix}
         p_{x}^{(i)} \\
         p_{y}^{(i)} \\
         p_{z}^{(i)} \\
        \end{bmatrix}
        = \begin{bmatrix*}[c]
          A^{(i)}_{n} & A^{(i)}_{H} & A^{(i)}_{V} \\
        \end{bmatrix*}
        \begin{bmatrix}
         p_{N}^{(i)} \\
         p_{H}^{(i)} \\3
         p_{V}^{(i)} \\
        \end{bmatrix},
\label{eq:f_vec_A_1}
\end{equation}
\end{linenomath*}
where
\begin{linenomath*}
\begin{equation}
        A_{n}
        = \begin{bmatrix*}[c]
          -\hat{n}_{1}^{\,(i)} \\
          -\hat{n}_{2}^{\,(i)} \\
          0 \\
        \end{bmatrix*}, \qquad
        A_{H}
        = \begin{bmatrix*}[c]
          \hat{t}_{H1}^{\,(i)} \\
          \hat{t}_{H2}^{\,(i)} \\
          0 \\
        \end{bmatrix*}, \qquad
        A_{V}
        = \begin{bmatrix*}[c]
          0 \\
          0 \\
         \hat{t}_{V3}^{\,(i)} \\
        \end{bmatrix*}.
\label{eq:f_vec_A_2}
\end{equation}
\end{linenomath*}

This linear elastic finite element model was implemented in FEniCS \citep{Fenics2015} using a tetrahedral mesh that was generated with the CUBIT software, using a CAD geometry of the buoy used in the experiment (Figure \ref{fig:gauge_locations}). The CAD files were provided by the University of Washington Applied Physics Laboratory.

\subsection{Bayesian inference}
\label{subsec:bayes_method}
We adopt a Bayesian approach to indirectly characterize the ice pressures $p_N(x), p_H(x), p_V(x)$ applied to a defined region on the exterior surface of the buoy from strain gauge measurements on the interior of the buoy.  Bayesian inference is a probabilistic framework that allows us to not only estimate the value of the pressures, but also to rigorously quantify uncertainty stemming from noisy observations and the inability of a few observations to completely constrain the pressures over the entire surface of the buoy.  Let $\Gamma_p\subset \Gamma$ denote the portion of buoy's exterior surface that lies between 23.5 and 83.5 cm from the top of the buoy.  Over this region, we want to infer the pressure function $p(x)$.  We start by treating the external pressures applied to the buoy and the observed strains as random variables, denoted by $p(x)=\left[p_N(x), p_H(x), p_V(x)\right]^T$ and $\varepsilon_{obs}$, respectively.   The components $p_N(x)$, $p_H(x)$ and $p_V(x)$ denote the stresses in the normal, horizontal, and vertical directions, as defined in Figure \ref{fig:force_diagram}.  

Our goal is to determine the conditional distribution of the pressures $p(x)$ given the strain observations $\varepsilon_{obs}$.  In general however, $p(x)$ lies in an infinite dimensional function space, which makes inference more challenging.  To overcome this, we instead look for pressure functions $p(x)$ that lie within the span of a linear finite element basis; the same basis used above to represent the load vector $f$.  This restriction allows us to characterize the pressure function $p(x)$ with a finite number of coefficients and thus apply standard Bayesian techniques.  \cite{stuart2010inverse} provides more information about working directly in the infinite dimensional setting.

Let $p$ (without the $(x)$) denote a vector containing nodal pressures. Our goal is then to characterize the posterior density $\pi(p | \varepsilon_{obs})$.  Bayes' rule allows us to write this density as the product of a prior density $\pi(p)$ and a likelihood function $\pi(\varepsilon_{obs} | p)$
\begin{equation}
\pi(p | \varepsilon_{obs}) \propto \pi( \varepsilon_{obs} | p) \pi(p),
\end{equation}
where the prior density models information known about the pressures before any observations are available, and the likelihood function provides a way of comparing model predictions with the observations.  More information is provided on each of these components below.

\subsubsection{Prior distribution}
To obtain a prior distribution over the finite dimensional vector $p$, we first consider the continuous pressure function $p(x)$, which is spatially varying field. It is natural to probabilistically describe the load function $p(x)$ as a random field.  In particular, we describe the prior distribution over the components $p_N(x),$ $p_H(x)$, and  $p_V(x)$ with independent Gaussian processes (see e.g., \citep{GPML2005}).  A Gaussian process defines a probability distribution over functions and can be interpreted as the infinite-dimensional analog of the more common multivariate Gaussian distribution.  Like a multivariate Gaussian distribution, which is completely defined by its mean vector and covariance matrix, a Gaussian process is completely defined by a mean function and a covariance kernel.    The covariance kernel describes the correlation between loads at two points $x$ and $x^\prime$ while the mean function describes the average pressure at a single point.  We use $\mu_N(x)$, $\mu_H(x)$, and $\mu_V(x)$ to denote the mean functions for each component and $k_N(x,x^\prime)$, $k_H(x,x^\prime)$, and $k_V(x,x^\prime)$ to denote the corresponding covariance kernels.  Together, these functions define Gaussian process distributions over each component
\begin{linenomath*}
\begin{eqnarray}
p_N(x) &\sim& GP\left(\mu_N(x), k_N(x,x^\prime) \right) \\
p_H(x) &\sim& GP\left(\mu_H(x), k_H(x,x^\prime) \right) \\
p_V(x) &\sim& GP\left(\mu_V(x), k_V(x,x^\prime) \right).
\end{eqnarray}
\end{linenomath*}
We further assume that all of the covariance kernels take the form
\begin{linenomath*}
\begin{equation}
k(x,x^\prime) = k_{12}\left(\tan^{-1}\left[\frac{x_2}{x_1}\right],\tan^{-1}\left[\frac{x^\prime_2}{x^\prime_1}\right]\right)k_3\left(x_3, x^\prime_3\right),
\end{equation}
\end{linenomath*}
where $x_i$ denotes component $i$ of the location $x$, $\tan^{-1}\left[\frac{x_2}{x_1}\right]$ is the angle around the buoy, $k_{12}$ is a 1d periodic covariance kernel defining the meridional (horizontal) correlation of $p(x)$, and $k_3$ is a Matern kernel with $\nu=3/2$ defining the vertical correlation.  The standard deviation (i.e., $\sqrt{k(x,x)}$) is set to $4.0$ MPa for $k_N$ and $0.5$ MPa for both $k_H$ and $k_V$.  These standard deviations were chosen to reflect \reviewchange{the} low probability that pressures would three times these values due to the mechanical properties of the ice.  The lengthscale used in all meridional kernels is $\frac{\pi}{20}$ while the lengthscale used in all vertical kernels is $50$ cm.  These were chosen based on the anticipated smoothness of the pressure fields.

As mentioned above, in order to use these Gaussian process descriptions with the finite element discretization described above, we need to discretize the pressure function $p(x)$.  To do this with our Gaussian process prior distribution, we evaluate the mean function at every FEM degree of freedom in $\Gamma_p$ and evaluate the covariance kernel for every pair of locations.  The result is a finite dimensional Gaussian distribution that is more convenient to work with than the infinite dimensional Gaussian process.  The density over the vector of nodal pressures $p$ is then given by
\begin{linenomath*}
\begin{equation}
\pi(p) = N(\mu_p,\Sigma_p)  = \frac{1}{\sqrt{\left|2\pi \Sigma_p\right| }} \exp\left[-\frac{1}{2} \left(p-\mu_p\right)^T \Sigma_p^{-1} \left(p-\mu_p\right)\right],
\label{eq:prior}
\end{equation}
\end{linenomath*}
where 
\begin{linenomath*}
\begin{equation}
p = \left[\begin{array}{c}p_N\\p_H\\p_V \end{array}\right], \qquad \mu_p =  \left[\begin{array}{c} \mu_N\\ \mu_H\\ \mu_V \end{array}\right], \qquad \Sigma_p = \left[\begin{array}{ccc} \Sigma_{N} & 0 & 0\\ 0 & \Sigma_{H} & 0\\ 0 & 0 & \Sigma_{V}\end{array}\right],
\end{equation}
\end{linenomath*}
and  $\mu_{N,i} = \mu_N(x^{(i)})$, $\Sigma_{N,ij} = k(x^{(i)}, x^{(j)})$, and $x^{(i)}$ denotes the location of mesh node $i$ in $\Omega_p$.  The Gaussian density $\pi(p)$ denotes the prior distribution over the external pressures.  

\subsubsection{Likelihood function}
The likelihood function describes the distribution of anticipated observations for a particular load $p$, i.e., What strains are likely to be observed for a particular load?  Our finite element model is a relatively accurate representation of the buoy.  We would therefore expect the modeled strain $\varepsilon_{mod}$, given by Eq.\eqref{eq:strain}--\eqref{eq:strain2}, to be close to the observed strain $\varepsilon_{obs}$ if the true pressures were used in the model.  With an accurate model, the difference between $\varepsilon_{mod}$ and $\varepsilon_{obs}$ is then mostly a result of noise in the observations.  We model this noise with a Gaussian random variable $z\sim N(0,\sigma_z^2 I)$ and assume the additive form
\begin{linenomath*}
\begin{equation}
	\varepsilon_{obs} = \varepsilon_{mod} + z .
\label{eq:strain_model}
\end{equation}
\end{linenomath*}
The likelihood function $\pi(\varepsilon_{obs} | p)$ is then a Gaussian density over the difference $\varepsilon_{obs} -\varepsilon_{mod}$.  In particular,
\begin{linenomath*}
\begin{equation}
\pi(\varepsilon_{obs} | p) = \left(2\pi\sigma_z^2\right)^{-N_z/2}\exp\left[-\frac{1}{\sigma_z^2} \left\|\varepsilon_{obs}- BK^{-1}DAp\right\|^2\right],
\label{eq:likelihood}
\end{equation}
\end{linenomath*}
where the finite element model defined in Eq.\eqref{eq:strain} and \eqref{eq:strain2} has been introduced to make the dependence on $p$ explicit.

\subsubsection{Posterior distribution}
Multiplying the Gaussian prior in Eq.\eqref{eq:prior} with the Gaussian likelihood in Eq.\eqref{eq:likelihood} yields the Bayesian posterior $\pi(p | \varepsilon_{obs})\propto \pi(\varepsilon_{obs} | p)\pi(p)$, which is also Gaussian and defined by a mean vector $\mu_{p|\varepsilon}$ and a covariance matrix $\Sigma_{p|\varepsilon}$.  To simplify notation, consider the prior predictive covariance
\begin{linenomath*}
\begin{equation}
\Sigma_{\varepsilon} = \left(BK^{-1}DA\right) \Sigma_p \left(BK^{-1}DA\right)^T + \sigma_z^2 I,
\end{equation}
\end{linenomath*}
and the Kalman gain
\begin{linenomath*}
\begin{equation}
G = \Sigma_p \left(BK^{-1}DA\right)^T \Sigma_{\varepsilon}^{-1}.
\end{equation}
\end{linenomath*}
The posterior mean and covariance are then given by
\begin{linenomath*}
\begin{eqnarray}
\mu_{p | \varepsilon} &= &\mu_p + G\left( \varepsilon_{obs} - \left(BK^{-1}DA\right)p\right)\\
\Sigma_{p|\varepsilon} &=& \Sigma_{p} - G \left(BK^{-1}DA\right) \Sigma_p.
\end{eqnarray}
\end{linenomath*}
The posterior mean $\mu_{p|\varepsilon}$ describes the most likely pressures while the posterior covariance $\Sigma_{p|\varepsilon}$ characterizes the remaining uncertainty.  Note that the block structure in $\Sigma_p$ and $A$ enable the posterior mean and covariance to be efficiently constructed without explicitly building or storing the full covariance $\Sigma_p$.  The Gaussianity of the prior and likelihood, combined with the use of a linear model, allows us to compute the posterior analytically rather than having to use computationally expensive sampling methods like Markov chain Monte Carlo.

\tikzstyle{block} = [rectangle, draw, text width=5em, text centered, minimum height=4em]
\tikzstyle{line} = [draw, -latex']

        


\section{Results}\label{sec:results}

\subsection{Inference Verification with Synthetic Strain Measurements}
\begin{figure*}[t]
	\centering
	\includegraphics[width=\textwidth]{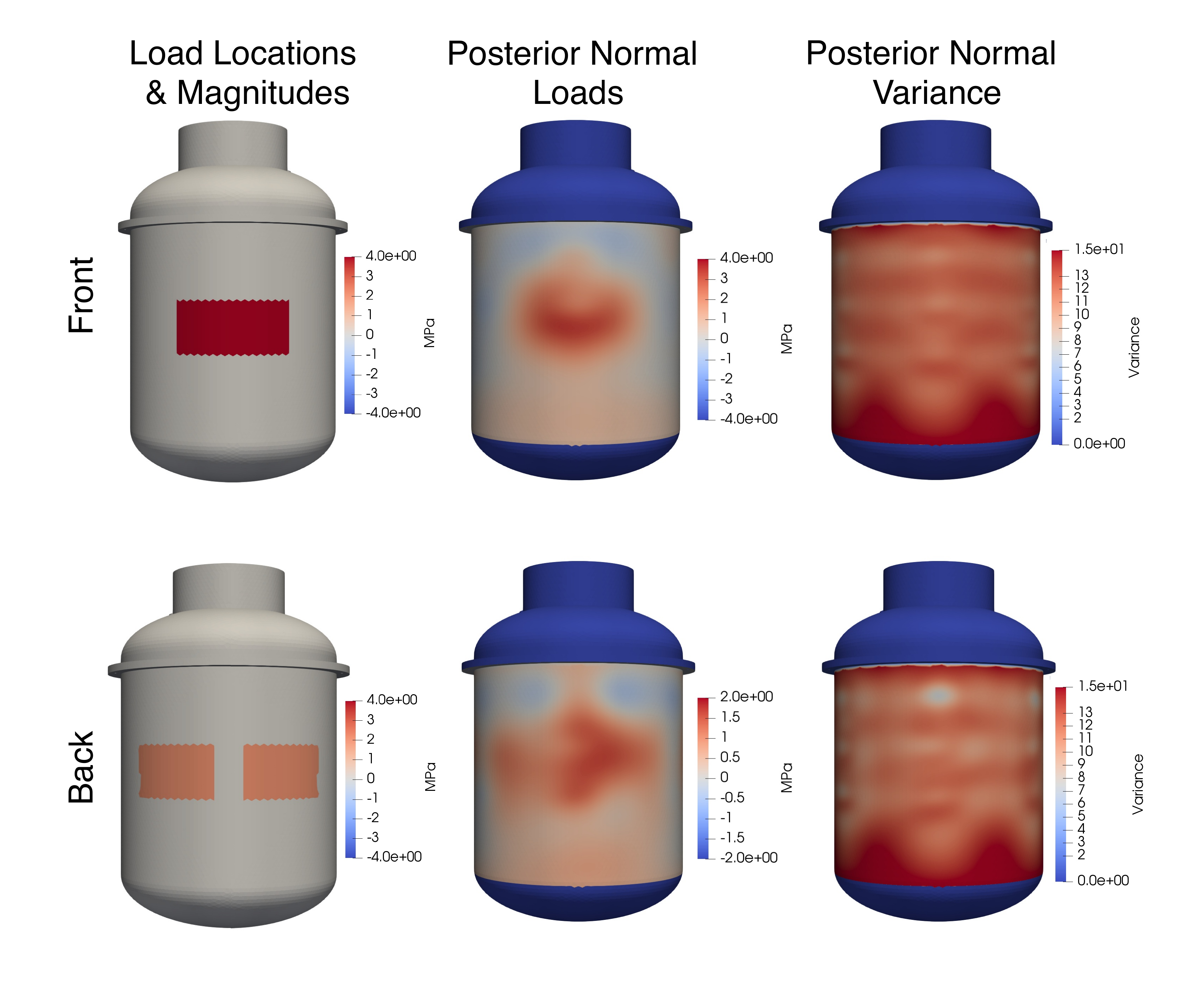}
	\caption{Synthetic load configuration and inference results.}
	\label{fig:Synthetic}
\end{figure*}

We used a set of synthetic strain measurements that resulted from a known load configuration to verify that the inference framework works as expected and provides a reasonable characterization of the applied pressures. The load configuration used for the verification was composed of 3 rectangular patches that were located with respect to the $x$-coordinate axis, which is aligned with the $\theta=0$ and $180^\circ$ direction, as seen in Fig. \ref{fig:Synthetic}. On the front side we applied a normal pressure of 4 MPa in a rectangular patch with dimensions 69.9 cm x 20.0 cm and on the back applied normal pressures of 2 MPa in two rectangular patches with dimensions 34.4 cm x 20.0 cm, placed an equal distance on either side of the $x$-axis. Fig. \ref{fig:Synthetic} shows the posterior mean of the normal load configurations. The inferred load pattern shows a region of elevated pressure centered at $\theta=0^\circ$ with a maximum pressure of 3.47 MPa. The pressures on the rear side of the buoy reach a maximum of 1.65 MPa. The spatial extents of both front and rear regions of high pressure compare well to the known applied load pattern. The spatial pattern of the variances is indicative of the added information that the strain gauges provide and the differences in the length scales for the horizontal and vertical directions in \reviewchange{the} Gaussian process used to represent the inferred load fields. The variance tends to be high as one moves farther away from a strain gauge with the decay in information being stronger in the vertical direction.  Interestingly, the prior lengthscale is actually shorter in the horizontal direction.  The slower increase of the posterior covariance in the horizontal direction is therefore an indication that this \reviewchange{strain} gauge configuration is more informative about horizontal variations in the applied pressures.

\begin{figure}[t]
	\centering
	\includegraphics[width=3.25in]{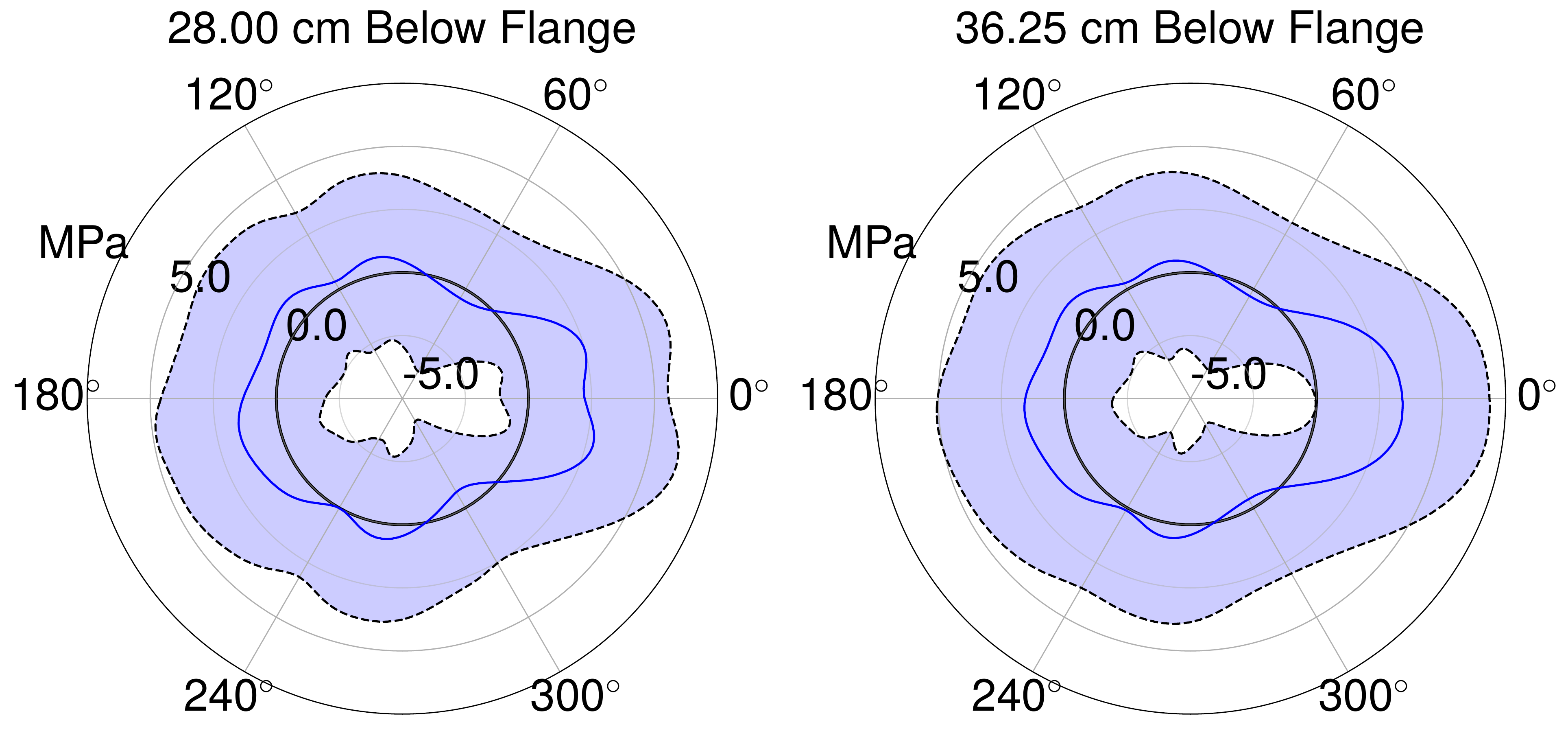}
	\caption{Detail of inferred normal load configuration at 28.0 and 36.25 cm below the flange.}
	\label{fig:Rose_synthetic}
\end{figure}

Figure \ref{fig:Rose_synthetic} provides a more detailed look at the inferred load configuration for horizontal slices at vertical locations of 28.0 and 35.25 cm below the buoy flange. There were aberrant negative loads, but these negative excursions are very small and do not exceed 0.87 MPa. The peak loads for the fore side of the buoy are 3.31 MPa and located at $\theta=357^\circ$ and on the aft side of the buoy are 1.56 MPa and located at $\theta=184^\circ$ with the load configuration exhibiting a very slight asymmetry. There were small lobes at $\theta=100^\circ$ and $260^\circ$, but these are relatively small in extent and magnitude.

\subsection{Experimental Strain Measurements}
\begin{figure*}
	\centering
	\includegraphics[width=\textwidth]{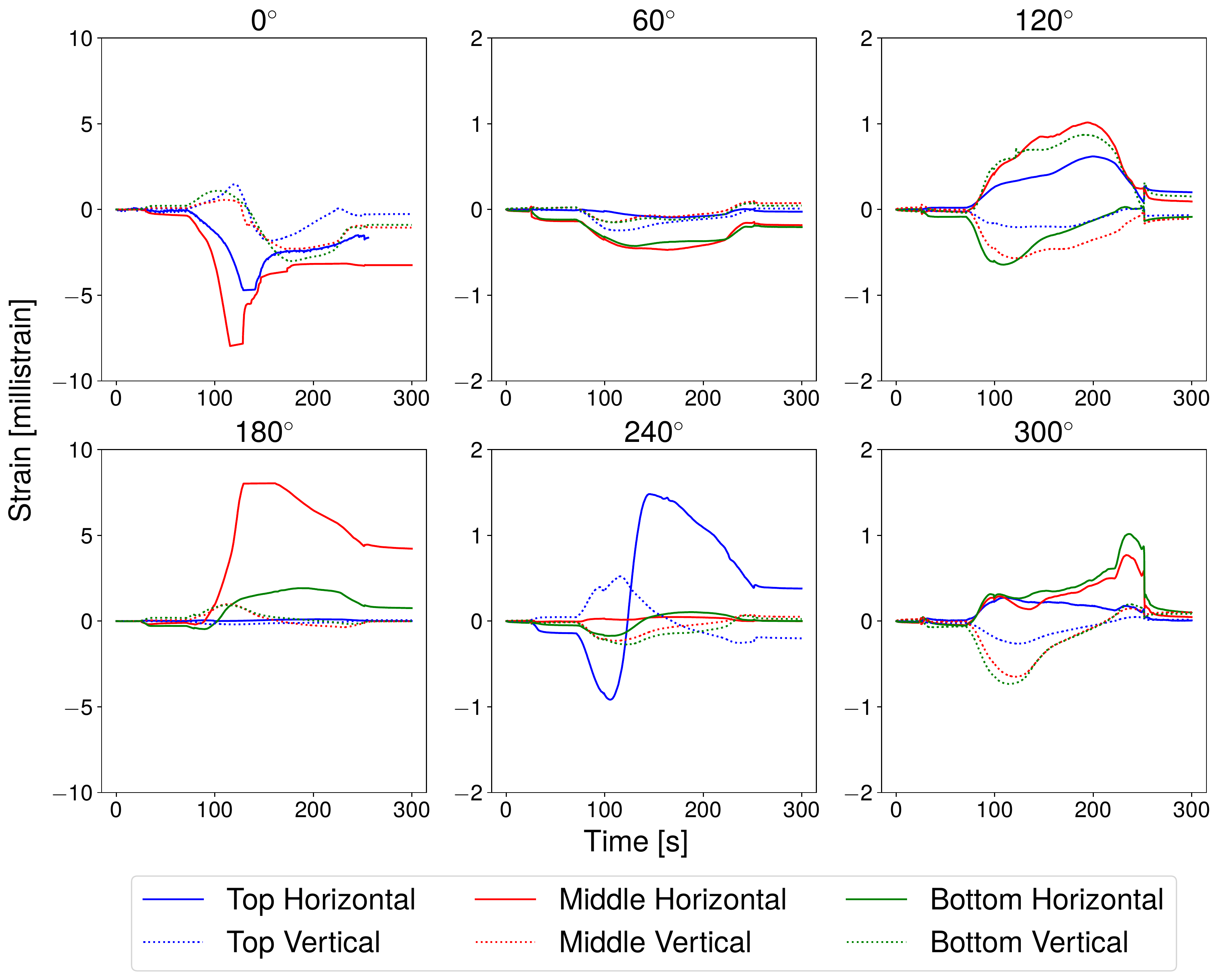}
	\caption{Horizontal and vertical strain gauge measurements from the compression experiment through time.  Each plot shows the top, middle, and bottom gauges for one orientation on the buoy (dashed lines are vertical strain gauges, and solid lines are the horizontal gauges).  Note the scale difference for the \ang{0} and \ang{180} plots versus the others.}
	\label{fig:Strain_Gauges}
\end{figure*}

As described in Section \ref{subsec:experiment}, we monitored the buoy's strain using foil gauges that were affixed to the interior surface of the buoy as an ice sheet was compressively loaded using a hydraulic ram.  Figure \ref{fig:Strain_Gauges}, shows how both the vertically and horizontally aligned strain measured at 18 locations on the buoy varied through the 300 second long experiment. As the ice was initially loaded, the strains quickly rise while the hydraulic ram applied approximately 7 MPa to the ice until the 73 second mark. During this initial period, the vertical strains measured at the locations normal to the direction of loading, $\theta=0$ and $180^\circ$, were all tensile as one would expect for a hollow body pinched in the middle. In contrast, the horizontal strains were mostly compressive.

For time period between 73-251 seconds, the hydraulic ram was set to its maximum output of 14 MPa through the duration of the experiment. During this timeframe, we observed a dramatic increase in strain rate for all the locations, and most locations reached a maximum value between 115 and 143 seconds. There were, however, a few exceptions to this strain trend. For example, the horizontal strains for the $\theta=120$ and $300^\circ$ locations, which face away and towards the hydraulic ram, respectively, continued to increase after 115 seconds, but still exhibited a reduction in strain rate when the other locations reached their maximum strain values. In general, the strains were either purely tensile or compressive through the duration of the experiment. However, at the $\theta=0$ and $240^\circ$ locations, the vertical strains transitioned from compression to tension, or vice versa, with the transitions occurring between 135 and 161 seconds. In addition, the horizontal gauge at $\theta=240^\circ$ transitioned from compression to tension at 124 seconds. 

The bottom horizontal strain gauge at \ang{0} stopped collecting data around 10 seconds, therefore was removed from the dataset used for the load inversion. 

\subsection{Inferred Loads}

\begin{figure}[t]
	\centering
	\includegraphics[width=3.25in]{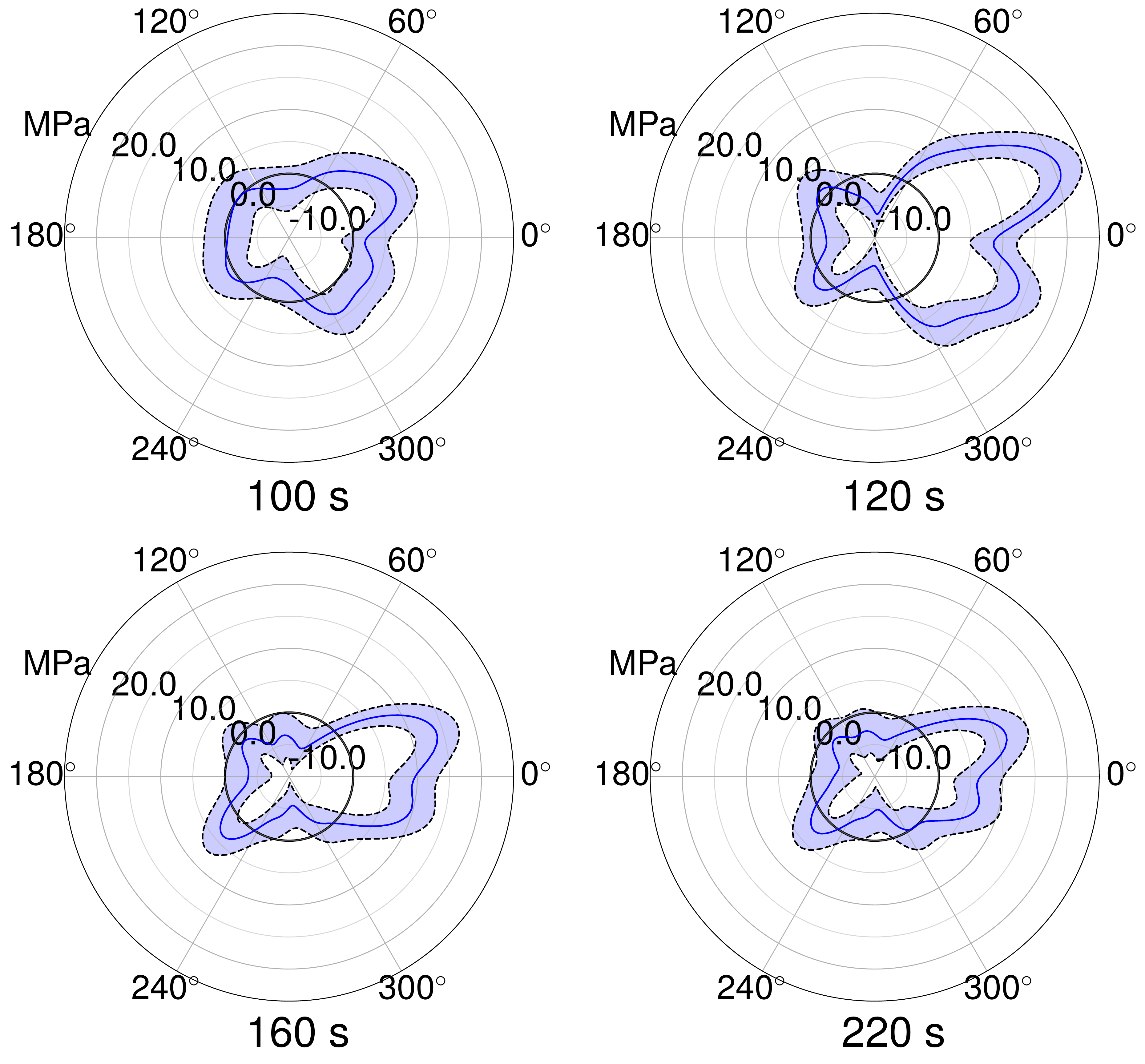}
	\caption{Rose plot showing the inferred normal load magnitudes (solid blue line) at times (a) 100, (b) 120, (c) 160, and (d) 220 seconds. The standard deviations associated with these loads is shown by the blue shaded region.  The thick black line is 0.0 MPa}
	\label{fig:loads}
\end{figure}

\begin{figure*}[t]
	\centering
	\includegraphics[width=\textwidth]{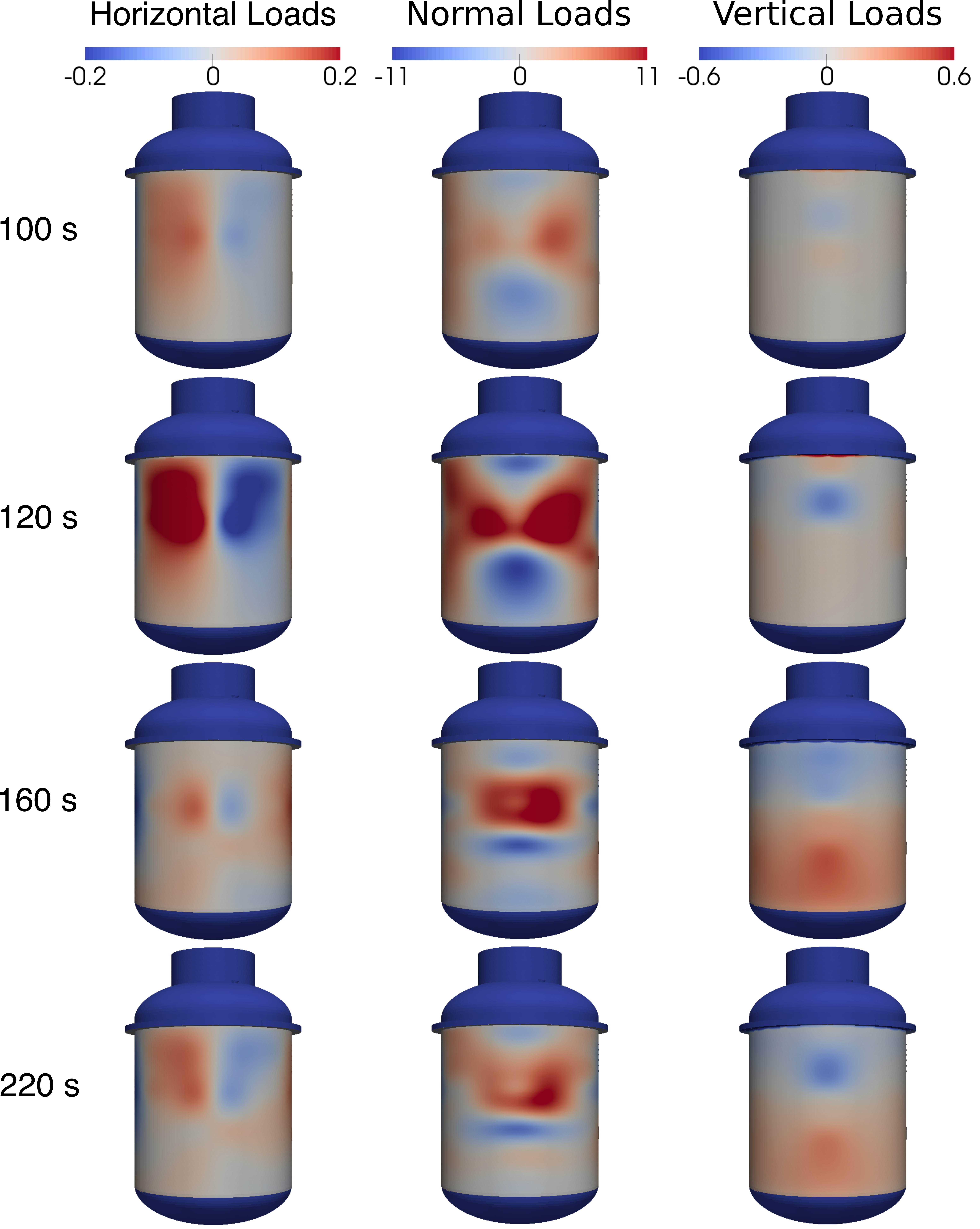}
	\caption{Inferred loads across the buoy looking down the direction of loading (\ang{0}) at times 100, 120, 160, and 220 seconds, respectively. The first column shows the horizontal loads, the middle column shows the normal loads, and the last column the vertical loads. The color bar units are MPa.}
	\label{fig:buoyLoads}
\end{figure*}

With the Bayesian inference framework introduced in Section \ref{subsec:bayes_method}, we used the measured strains of the buoy interior wall to infer the loads on the buoy's exterior surface. We performed this inference on a subregion of the buoy surface, $\Gamma_p$, that bounded the region that the ice was expected to be in contact with the buoy.  We independently processed each observation to obtain a time series of applied pressures, as shown in Figure \ref{fig:loads} and Figure \ref{fig:buoyLoads}. The inference results are presented as stress fields in the buoy coordinate system, with surface normal direction (positive pointing towards the buoy interior), the horizontal direction (positive pointing clockwise) and the vertical direction (positive pointing up).

\begin{figure}
	\centering
	\includegraphics[width=2.5in]{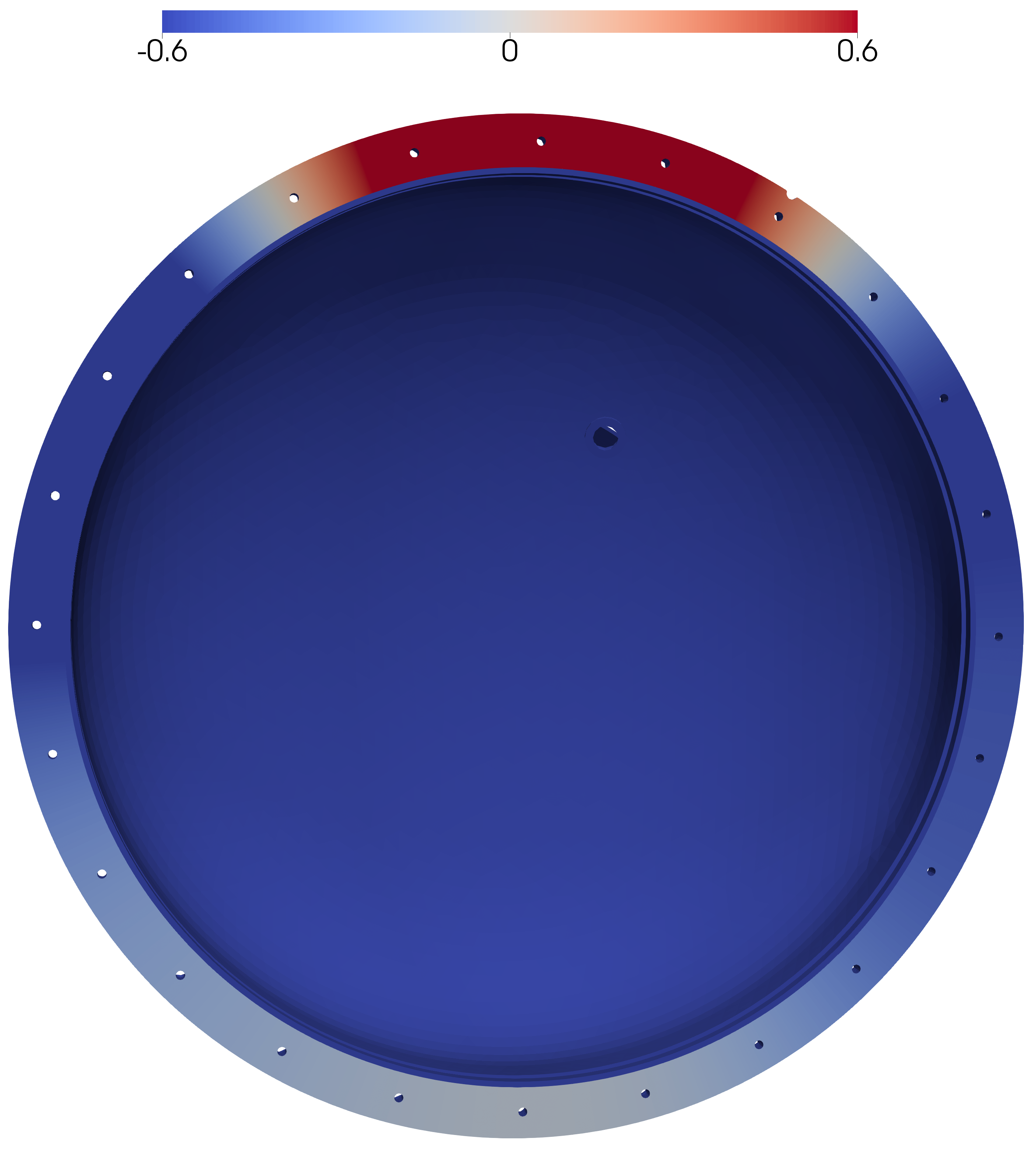}
	\caption{Detailed view of vertical pressure on the flange at 120s looking from below.  Positive quantities are pointing into the page.  There is a clear lift corresponding to the visible flange-ice contact in image B of Figure \ref{fig:Experiment_Pics}. Units are in MPa.}
	\label{fig:verticalDetail}
\end{figure}

For the initial 73 seconds, the stress levels on the buoy are very low, which is to be expected since the strains were also small during this time interval. When the hydraulic ram output is increased from 7 MPa to 14 MPa, the magnitude of the inferred stresses begin to increase until the horizontal, vertical, and normal stresses reach maxima of 2.7, 1.1, and 20.8 MPa, respectively, near a time of 125 seconds into the experiment. The posterior mean stress fields at 100 and 125 seconds exhibit lobes near \ang{0} that are consistent with the ice squeezing the buoy at this location and the formation of a crack at \ang{0}. The horizontally aligned stress shows a line of zero stress at $\theta=0^\circ$ with a positive and negative region on either side of this zero stress line. This stress divergent pattern about the $0^\circ$ line on the buoy is further evidence that a crack had formed at this location. In contrast, the vertical stress pattern shows a zero stress line aligned horizontally with a negative stress (pushing down) above and a positive stress (pushing up) below this line, i.e., tangential stresses that are converging toward this line. The largest vertical stresses are found along the bottom of the buoy's flange, as shown at time 120s in Figure \ref{fig:verticalDetail}.  Between 100 and 125 seconds, the vertical stresses on the flange near \ang{0} are positive indicating that ice was pushing up on flange during these times.  As Figure \ref{fig:Experiment_Pics} illustrates, this pressure on the flange agrees with visible observations of the buoy being lifted by the flange during the experiment. 

Figure \ref{fig:loads} provides a more detailed look at the inferred normal stresses along the middle ring of strain gauges (51.2 cm from the top of the buoy) at 100, 120, 160, and 220 seconds. We see the large magnitude lobes pressure on either side of \ang{0} and how the asymmetry increases with time favoring the sextant between \ang{0} and \ang{60}. There are region of negative stress throughout the experiment that we attribute to an unloading from a prestress in the buoy due to the freezing in process, therefore we suspect that there is an unknown biases in the strain measurements that we were unable to determine. The maximum inferred pressure in the lobe located in the sextant between \ang{0} and \ang{60} was 20.5 MPa and the maximum inferred pressure in the lobe located in the sextant located between \ang{320} and \ang{0} was 13.2 MPa .

\subsection{Posterior predictive strains and displacement}

As a sanity check, we calculate the posterior predictive mean strain at the observation locations, i.e., the strain $\varepsilon_{post}$, by using the posterior mean pressures as the boundary conditions for the FEM forward model. We compared these strain values against the actual observed strain values, $\varepsilon_{obs}$, as a verification that the posterior loads were matching the observations. The largest difference between the mean posterior predictive and the observed strains was $3.03\times10^{-11}$, and occurred at 145 seconds, which corresponds to shortly after the time of highest observed strain.

\begin{figure}
	\centering
	\includegraphics[width=3.25in]{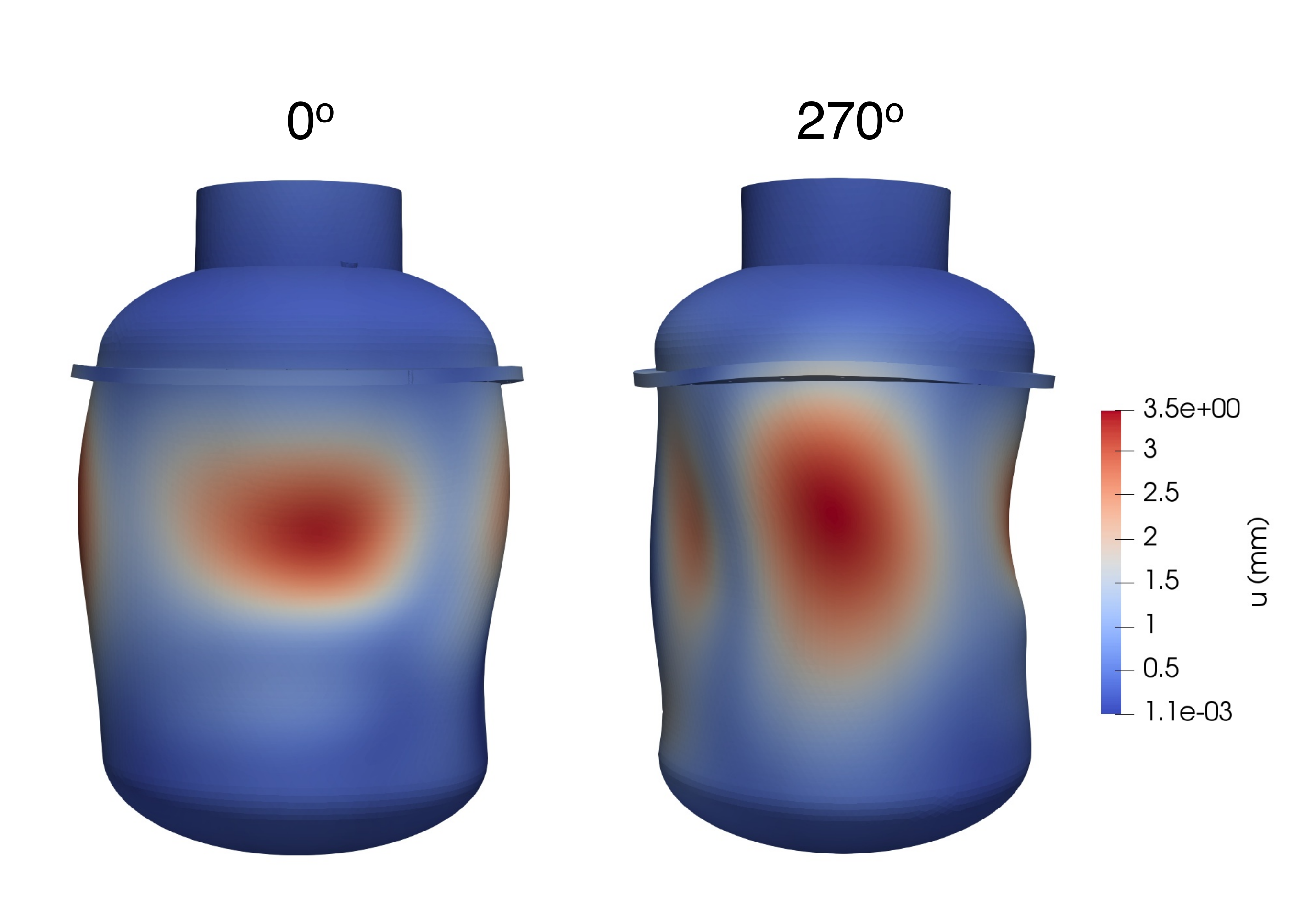}
	\caption{Calculated displacements at 200 seconds using the inferred load results.  The buoy is warped by a scaling factor of 2.0, and is show oriented along the \ang{0} and \ang{270} axes}
	\label{fig:fullField}
\end{figure}

The inferred loads were also used to calculate the expected posterior displacement of the buoy surface.  The largest displacements were at 175 seconds, along the \ang{0} axis ($17$ mm), where the buoy deformed inward.  In addition to displacement in the direction of loading, the buoy bowed outward along the \ang{90} and \ang{270} axes ($12$ mm and $14$ mm, respectively) in response to axial compression. The displacement fields align well with the distribution of the normal loads, where highly compressive loads are found in the same areas on the buoy where inward deformation occurred. The displacement results at 200 seconds are shown in Figure \ref{fig:fullField} to illustrate the final deformation state of the buoy.  These results help to shape our understanding of the complex loading scenario that caused the observed strains.  

\subsection{Visual observations of experiment}

\begin{figure*}[t]
	\centering
	\includegraphics[width=\textwidth]{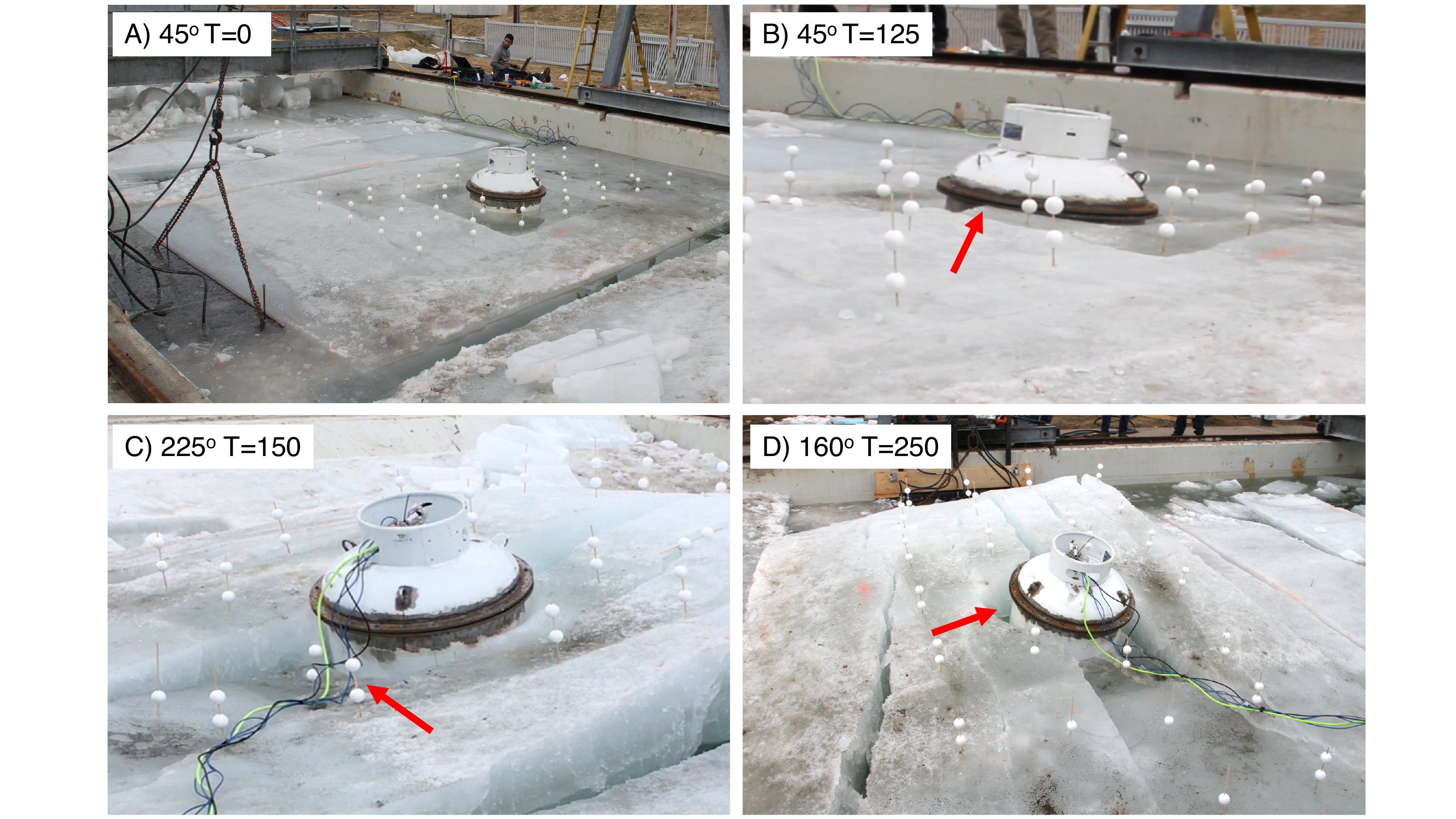}
	\caption{ Images from the physical experiment showing features from varying viewing angles, relative to the \ang{0} direction of loading. Red arrows indicate features highlighted for comparison with inferred loads. A) Initial setup before any deformation; B) When ice begins to catch on the flange and raise the buoy; C) Ice and buoy are out of contact on the backside of the buoy after large chunk of ice falls away; D) Large gaps between the buoy and original ice contact surfaces.}
	\label{fig:Experiment_Pics}
\end{figure*}

To evaluate how well the inference framework worked for this experiment, we qualitatively correlate the inferred load configurations with key events during the physical experiment. Initially, the buoy was frozen into the ice and had full contact between the buoy and the ice through the ice thickness. As the ice was compressively loaded, the buoy acted as an inclusion in the ice and induced localized ice cracking and deformation around the buoy. In some locations, the buoy decoupled from the ice and transferred the load to highly localized regions on the buoy. At approximately 100 seconds, a large crack formed in front of the buoy along the direction of loading along with two cracks off the backside of the buoy; one along the \ang{180} axis and another emanating from the edge of the buoy near \ang{240}. These two cracks on the backside separated a small segment of ice from the rest of the ice sheet, as observed in images C and D of Figure \ref{fig:Experiment_Pics}. The normal loads on the front of the buoy presented in Fig. \ref{fig:buoyLoads} showing compression in two lobes corroborates the presence of a crack in the loading direction. The divergent horizontal loads in the \ang{0} direction also indicate that the two sides of this crack were spreading apart as they pressed against the buoy. This is supported by observations that the crack became wider as the experiment progressed.  

The ice did not remain planar throughout the experiment and started to buckle at 83 seconds, with some sections pushing up nearly 1 meter above the initial elevation. As it buckled, the ice sheet lifted the buoy unevenly. The uneven lifting action was partly a result of the major crack that formed through the middle (at \ang{0}) of the ice sheet.  Wlipping between the ice and buoy was also observed until the ice engaged with the buoy flange (see image B in Fig \ref{fig:Experiment_Pics}). The inferred load configurations show large vertical loads on the flange and are highest  upward load on the flange at 120 seconds in Figures \ref{fig:buoyLoads} and \ref{fig:verticalDetail} consistent with the observations. As described previously, the ice load is higher in the sextant between \ang{0} and \ang{60} matching the visual observations in the experiment where ice exhibiting more engagement with this sextant.

Another key observation from the experiment was that the small separated section of ice on the back edge of the buoy \reviewchange{rose} with the rest of the ice sheet as it buckled, which is an indicator that the a significant portion of the buoy's back side, i.e., between \ang{90} and \ang{270}, was not in contact with ice for a large portion of the test (image C in Figure \ref{fig:Experiment_Pics}), which may explain why the normal loads were inferred to be either small or even negative (see Fig. \ref{fig:loads}). \reviewchange{Note that by inferring the loads acting on the buoy, we are not explicitly modeling the ice-buoy contact forces.  A loss of contact manifests as a decrease in the load from the initial state of the buoy (i.e., negative load in Fig. \ref{fig:loads}).  The posterior probability of a non-negative load could therefore be used to characterize the probability of ice-buoy contact at any location on the buoy.} These results and corresponding observations illustrate how our inference framework could qualitatively capture the dynamic and complex loading conditions within the ice sheet that evolved during the experiment. 

\section{Discussion}\label{sec:discussion}


The good qualitative comparison between the ice behavior and inferred pressures shows promise for our approach.  However, there are certainly limitations as well.  Certain aspects of the inferred load distribution, such as the high normal loads on the buoy in the direction of loading, are easily explained with observations from the uniaxial compression test. But, other features of the inferred loads are more difficult to justify using the visual observations, such as the negative stress values on the buoy's backside. Additional experiments with more careful control of the loading conditions will be a better validation exercise for the approach presented in this paper. Another limitation is the small number of observation points ($N_{obs}=33$) relative to the number of degrees of freedom ($N_{dofs}=35,073$), which makes the inference problem inherently more challenging because we are trying to predict stress fields with a very sparse dataset. Of course, more observation points would improve results, but may not be feasible. \reviewchange{Preliminary tests leveraging Bayesian experimental design techniques have shown promise in optimizing strain gauge locations, but we leave a rigorous study of this approach to future work.}

As mentioned previously, we assumed linear elasticity for the forward model to reduce computational complexity, but the buoy did plastically deform breaking this assumption and possibly causing the large inferred pressures. Therefore, a structural model for the buoy that can accommodate plastic deformation should improve the inference results, but it should be noted that a nonlinear model poses added computational challenges \reviewchange{since the posterior would no longer be Gaussian and a sampling strategy such as Markov chain Monte Carlo (MCMC) would be required.   High dimensional MCMC for random fields is an active area of research (e.g., \cite{cui2016dimension, cotter2013mcmc}) and could be useful for future work in this setting.  However, we expect the typical operating regime of deployed buoy systems to be within elastic limits even though this preliminary experiment resulted in plastic deformation.}

\reviewchange{Being an indirect way of measuring ice stress, our approach will likely be less accurate than existing sensors, such as vibrating wire stress gauges, that are explicitly designed for high resolution stress measurements.  However,  our approach has the potential to turn any existing ice buoy system into a stress sensor and could therefore lead to much wider temporal and spatial coverage.}

\section{Conclusions}\label{sec:conclusion}

We have demonstrated a Bayesian inference framework that infers external load conditions on a buoy using a network of  gauges placed on the interior surface of a steel cylindrical buoy. Using the inferred loads as inputs to a structural model of the buoy, we verified the inferred results by comparing the modeled strain against the observed strains. The  Bayesian inference approach presented in this paper provides a robust framework for estimating the quantity of interest, in our case the most likely load configuration, but also the associated uncertainty, which helps assess the quality of the inference results.

Distributed observation of internal ice stress within the Arctic ice pack is crucial for improving the understanding of sea ice dynamics. To date, observations of sea ice internal stress are sparse or non-existent and certainly are not part of an ice monitoring system. \reviewchange{We believe the framework described here could allow existing buoy systems to be used as a network of stress sensors that can help capture the highly variable stress conditions in sea ice.} Continuous, in situ internal stress measurements will allow us to observe how these stresses develop, evolve, and propagate through the ice pack leading to improvements in our understanding of the mechanics behind local processes like ridging and lead formation. This information will be crucial for improving sea ice models and providing operational guidance in the region. The physical character and dynamic behavior of Arctic sea ice will continue to change in response to the warming climate and instrumentation that can provide new insight into the mechanical behavior sea ice will be essential. We believe the Bayesian inference framework presented in this paper is a promising step in that direction.

\section{Acknowledgements}\label{sec:acknowledgements}
We would like to thank David Dyer, Jestoni Orcejola, and Dr. Sarah Webster from the University of Washington Applied Physics Laboratory for providing the buoy used in the experiment and the strain gauge data.  We would also like to thank Dr. Andrew Davis for taking the time to provide constructive feedback for this manuscript. 

\textit{This study was supported by the Office of Naval Research Arctic and Global Prediction program (code 32) under award N0001418MP00501.}

\bibliographystyle{asmejour} 
\bibliography{BuoyInversion}

\end{document}